# Size-selective optical printing of silicon nanoparticles through their dipolar magnetic resonance

Cecilia Zaza,[1,2] Ianina L. Violi,[1*] Julián Gargiulo,[3] Germán Chiarelli,[1,2] Ludmilla Schumacher,[4] Jurij Jakobi,[5] Jorge Olmos,[6] Emiliano Cortes,[3,7] Matthias König,[4] Stephan Barcikowski,[5] Sebastian Schlücker,[4] Juan José Saenz,[6,8] Stefan A Maier,[3,7] Fernando D. Stefani[1,2*]

1. Centro de Investigaciones en Bionanociencias (CIBION), Consejo Nacional de Investigaciones Científicas y Técnicas (CONICET), Godoy Cruz 2390, CABA, Argentina
2. Departamento de Física, Facultad de Ciencias Exactas y Naturales, Universidad de Buenos Aires, Güiraldes 2620, CABA, Argentina
3. The Blackett Laboratory, Department of Physics, Imperial College London, London SW7 2AZ, United Kingdom
4. Physical Chemistry I, Department of Chemistry and Center for Nanointegration Duisburg-Essen (CENIDE), University of Duisburg-Essen, Germany
5. Technical Chemistry I, Department of Chemistry and Center for Nanointegration Duisburg-Essen (CENIDE), University of Duisburg-Essen, Germany
6. Donostia International Physics Center (DIPC), 20018, San Sebastián, Spain
7. Chair in Hybrid Nanosystems, Nanoinstitute Munich, Faculty of Physics, Ludwig-Maximilians-Universität München, 80539 München, Germany
8. IKERBASQUE, Basque Foundation for Science, 48013 Bilbao, Spain

ABSTRACT

Silicon nanoparticles possess unique size-dependent optical properties due to their strong electric and magnetic resonances in the visible range. However, their widespread application has been limited, in comparison to other (e.g. metallic) nanoparticles, because their preparation on monodisperse colloids remains challenging. Exploiting the unique properties of Si nanoparticles in nano- and micro-devices calls for methods able to sort and organize them from a colloidal suspension onto specific positions of solid substrates with nanometric precision. Here, we demonstrate that surfactant-free Silicon nanoparticles of a predefined and narrow ($\sigma$ < 10 nm) size range can be selectively immobilized on a substrate by optical printing from a polydisperse colloidal suspension. The size selectivity is based on differential optical forces that can be applied on nanoparticles of different sizes by tuning the light wavelength to the size-dependent magnetic dipolar resonance of the nanoparticles.

TEXT

The theoretical and experimental advances in plasmonics have led to remarkable control in squeezing, enhancing and manipulating optical electric fields on the nanoscale.[1–3] More recently, a new research front has emerged focusing efforts on understanding and exploiting optical magnetic resonances of high refractive index (HRI) nanoparticles (NPs). Initially conceived to control the magnetic component of light through optically induced Mie resonances, HRI NPs have found further applications due to their reduced dissipative losses in comparison to metals and to the possibility of confining both electric and magnetic fields.[4–9]

These features promoted the investigation of dielectric optical nano-antennas made of HRI NPs as the magnetic homologs of metallic (plasmonic) nano-antennas.[10–14] Among the many different HRI dielectrics, silicon (Si) NPs and nano-antennas have been the most studied.[12,15,16]

Up to now, the great majority of experimental work has been done with nanostructures fabricated by means of top-down methods, such as electron beam lithography, which are highly developed for Si[12,17]. Top-down methods provide tight control over shape and size of Si nanostructures, but have a number of limitations for certain applications. They are intrinsically serial (slow, low throughput) and two-dimensional (e.g. nanospheres cannot be produced). Also the combination of different materials is highly demanding. Typically, these limitations are eluded with colloidal chemistry preparation, which is highly versatile for various materials, but remains challenging for HRI materials, and Si in particular. Attempts to obtain monodisperse colloids of Si NPs by chemical methods turned out to be successful only for rather large NP sizes. As such, Shi and co-workers recently synthesized monodisperse spherical Si NPs of ~ 425 nm diameter soluble in toluene.[18] This has motivated the exploration of alternative preparation methods of Si NPs. Femtosecond laser irradiation of a Si thin film target generates nano droplets of molten Si that can be transferred through a gas phase to a substrate, and produce spherical Si NPs in the size range of 150-250 nm.[14,19–21] Very small Si spheres of ~ 5 nm diameter were recently obtained using molten salts or laser ablation in organic solvents.[22–25] Chemical vapor deposition can deliver highly polydispersed Si spheres in the range of 250-5000 nm. Pulsed laser ablation in liquid solvents (LAL) is a low-cost, fast and simple method for obtaining surfactant-free colloids[26]. Kabashin et al. showed that Si colloids with narrow size distribution may be prepared via femtosecond laser irradiation of microparticles suspended in liquid, and proved the biocompatibility as well as the biodegradability of the partially oxidized Si NPs in vivo[27]. In general, the laser-based colloid synthesis method gives access to mostly spherical Si NPs in the range of 5-200 nm, depending on the pulse duration, repetition rate, wavelength, and solvent.[23,26,28,29] This range of sizes is particularly attractive for optical applications as magnetic and electric dipolar modes lie in the visible range.[30] However, these often polydisperse solutions present broad and featureless optical spectra as a result of the overlapped responses of all the different sizes of Si NPs, with magnetic and electric dipolar resonances covering a large bandwidth of the visible spectrum. Consequently, chemical and physical preparation methods of Si NP colloids as well as other applications such as sensors[16] and drug delivery[31] would highly benefit from any type of size-purification or sorting technique.

NPs with size-dependent optical properties can be sorted according to their size using optical forces.[32,33] In particular, the optical printing technique uses optical forces to capture NPs one by one from a liquid suspension and deliver them onto a solid substrate with size selectivity, high positional accuracy and a versatility of pattern design.[34–39] Meanwhile, the physical mechanisms involved in optical printing are well understood.[35] Although potentially applicable to any colloidal NP, optical printing of colloidal NPs has been only used for metal plasmonic nanoparticles in wide range of different sizes and shapes.[40–44]

Here, we demonstrate the size-selective optical printing of Si NPs from a polydisperse colloidal suspension produced by LAL of a Si target. First, the possibility of using a focused monochromatic beam to selectively push spherical NPs of a particular size is studied theoretically. Then, Si NPs were printed on glass substrates using visible lasers. It is shown that the size of the printed NPs can be chosen by tuning the wavelength of the laser beam to the dipolar magnetic resonances. Furthermore, correlated electron microscopy and single particle scattering spectroscopy on the printed NPs reveals a magnetic dipolar resonance determined by the NPs size. This work settles the basis for the optical printing of nano devices based on HRI colloidal NPs, where the optical properties, size and position of each individual component can be controlled.

EXPERIMENTAL METHODS

**NPs Synthesis and characterization**

Si NPs were generated via LAL[26] in acetone (99.9 %, Sigma Aldrich), using a picosecond laser (Ekspla, Atlantic 532) with a wavelength of 1064 nm, a repetition rate of 100 kHz, a pulse energy of 160 μJ, and a pulse duration of 10 ps. The Si NPs collected in acetone were transferred to ultra-pure water through centrifugation and redispersion. The obtained Si NPs are negatively charged due to the partial oxidation and the presence of Si-O- surface groups.[23,26] NPs characterization (shape, size, size distribution and crystallography) was performed by transmission electron microscopy (TEM - Zeiss, EM910). The polycrystalline structure of Si NPs was demonstrated by selected area electron diffraction (SAED) measurements. Size distribution was determined from the analysis of TEM images of 468 particles.

**Optical printing**

Substrates were surface functionalized using a layer-by-layer deposition of polyelectrolytes following the same protocol as in previous works.[35,41] Briefly, glass coverslips were first cleaned through the following steps: sonication for 10 minutes in Hellmanex (0.2% v/v), thorough rinsing with MilliQ water, rinsing with acetone, drying at 80 °C, and plasma cleaning for 3 minutes. Then, the substrate surface was modified by immersion for 15 minutes in a solution of polydiallyldimethylammonium chloride (PDDA) Mw = 400.000-500.000 (1 mg/ml in 0.5 M NaCl), rinsing with MilliQ water, immersion for 15 minutes in a solution of sodium polystyrene sulfonate (PSS) Mw = 70.000 (1mg/ml in 0.5 M NaCl), rinsing with MilliQ water. The prepared samples were stored in MilliQ water until use.

Dark field images were acquired using a digital colour camera. Single nanoparticle scattering spectra were obtained with a spectrometer (Shamrock 500i, Andor) based on 150 lines/mm grating and a EMCCD camera (Andor Ixon EM+ 885). Field-emission scanning electron microscopy (FE-SEM) images were obtained with a ZEISS LEO 982 GEMINI electron microscope in the secondary electron mode, using an in-lens detector to improve resolution.

RESULTS AND DISCUSSION

Figure 1A shows a representative transmission electron microscopy (TEM) image of the obtained polydisperse colloid. The size distribution is shown in Figure 1B; the NP radii ranges from 10 to 90 nm. In addition, electron diffraction patterns of single NPs (e.g. Figure 1A – inset) indicate that the NPs are polycrystalline. Under the dark-field optical microscope, each NP is observed as a bright and colorful spot, with different colors depending on the size (Figure 1C). Hence, the bulk colloid extinction spectrum shows a broad peak in the visible range (Figure 1D).

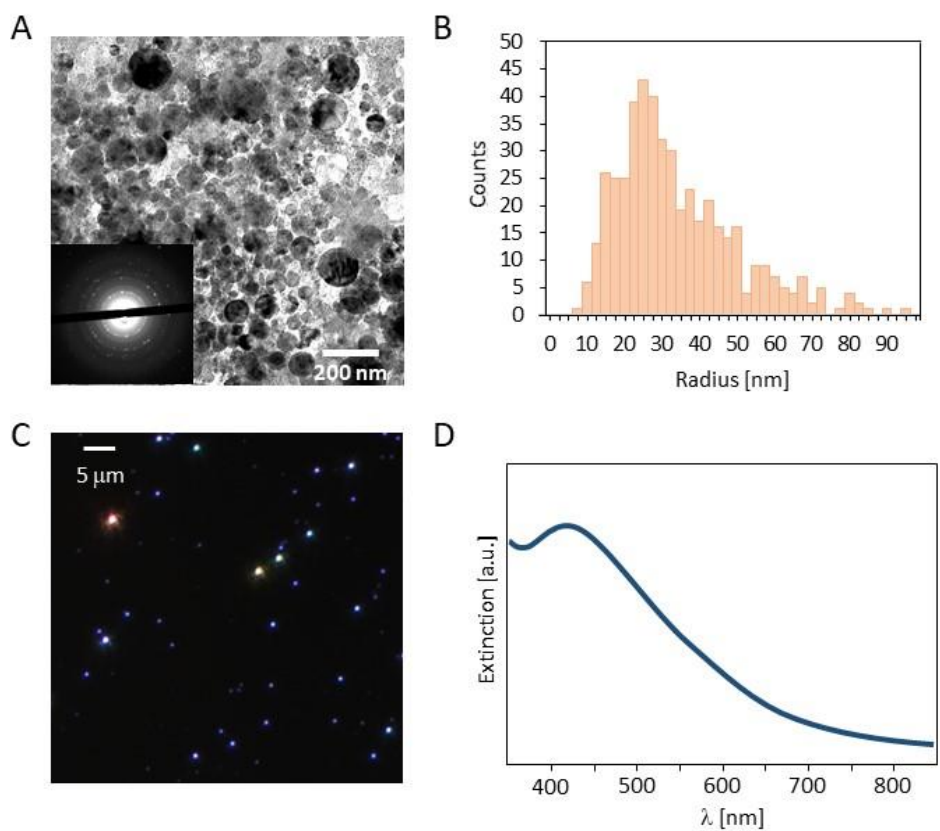

**Figure 1.** Characterization of the surfactant-free Si NP colloid obtained by picosecond-pulsed laser ablation in liquid phase (LAL). (A) TEM image; inset: example diffraction pattern of a single NP. (B) Size distribution of the NPs obtained from TEM images. (C) Dark field micrograph of Si NPs. (D) Extinction spectra of the Si NP colloid.

Silicon NPs have been manipulated optically in liquid using optical tweezers.[45] The potential of optical printing for sorting Si NPs from a polydisperse colloid was first evaluated theoretically. Recently, Shilkin et al[46] have shown theoretically that the radiation pressure exerted by a monochromatic plane wave on a HRI particle presents well defined resonances as a function of the particle size which could, in principle, be used in optical sorting of Si NPs. In our experiments, the Si NPs interact with a tightly focused beam, where the generated radiation pressure presents a spectral dependency that is quite different from the plane wave case. While a plane wave can efficiently excite high order multipoles, a tightly focused beam excites preferentially the lowest multipolar resonances.[47,48] We consider the most general case of a "dipolar" electric and magnetic focused beam with well-defined helicity $\sigma$ (i.e., a beam whose field expansion around the focus can be described by the first dipolar vector spherical harmonics) was considered

$$\boldsymbol{E}_\sigma = E_0 [i\sqrt{6\pi}] \left[ j_1(kr) \boldsymbol{X}_{1\sigma} + \sigma \frac{\nabla \times [j_1(kr) \boldsymbol{X}_{1\sigma}]}{k} \right] \quad (1)$$

where $E_0$ is the field intensity, $j_1(kr)$ is the $l=1$ spherical Bessel function and $\boldsymbol{X}_{11}$ the vector spherical harmonic (as defined by Jackson).[49] This dipolar beam is identical to the first term of the expansion of a circularly polarized plane wave around the origin. A linearly polarized dipolar beam can be obtained as a linear combination of fields with opposite helicities: $\boldsymbol{E}_x = (\boldsymbol{E}_- + \boldsymbol{E}_+)/\sqrt{2}$ (see details in the Supporting Information –SI-). Formulae for computing optical forces on spherical particles located in more general laser beams can be found in Gouesbet and Lock works.[50,51] When the sphere center is located at the origin, i.e., in the focus, the beam can only excite the dipolar electric and magnetic modes, even for large particles. Under this condition, the force can be shown to be independent of the beam polarization and it is given by (see SI for details)

$$\mathbf{F} = \mathbf{z}\frac{\epsilon_0 n_h^2}{2}|E_0|^2 \pi R^2 Q_{pr} \quad ; \quad Q_{pr} = \frac{2}{(kR)^2}\left[3 \ \text{Re}\left(\frac{a_1+b_1}{2} - a_1 b_1^*\right)\right] \quad (2)$$

where $R$ is the radius of the NP, $Q_{pr}$ is the radiation pressure efficiency, $n_h$ refractive index of the medium, and $a_1$ and $b_1$ are the resonant electric and magnetic coefficients, which are related to the electric and magnetic polarizabilities as $\alpha_E = i(6\pi/k^3)a_1$ and $\alpha_M = i(6\pi/k^3)b_1$, respectively. This simple expression for an isolated, free standing Si NP can be obtained from the so-called Generalized Lorentz-Mie theory or directly through the known results of a dipolar excited particle in an arbitrary exciting field[52]. In this particular case, the influence of the substrate on the Si NPs is not relevant regarding the magnetic resonance wavelength.[53] Interestingly, for the same field intensity $|E_0|^2$ at the particle position, the force induced by the focused beam is smaller than the one produced by a plane wave (even for dipolar small particles). This can be understood as a consequence of the curvature of the wave fronts in the focused beam, which reduces the projection of the incoming momentum along the beam axis (Figure 2A).[50] The radiation pressure efficiency given by Eq. (2) is a function of the size parameter $kR = 2\pi n_h R/\lambda$ and the relative refractive index $n_{Si}/n_h$ and presents a maximum ($Q_{pr} \sim 2.5$) at $kR_0 \sim 1$ as shown in Figure 2B and 2C (see also Figs S2-S4 in the SI). For each radius, there is one dominating resonance that can be attributed to the magnetic dipolar mode. Reciprocally, a monochromatic focused beam is expected to selectively push NPs of a small range of sizes. Furthermore, the size range can be tuned by changing the beam wavelength, as shown in Figure 2C.

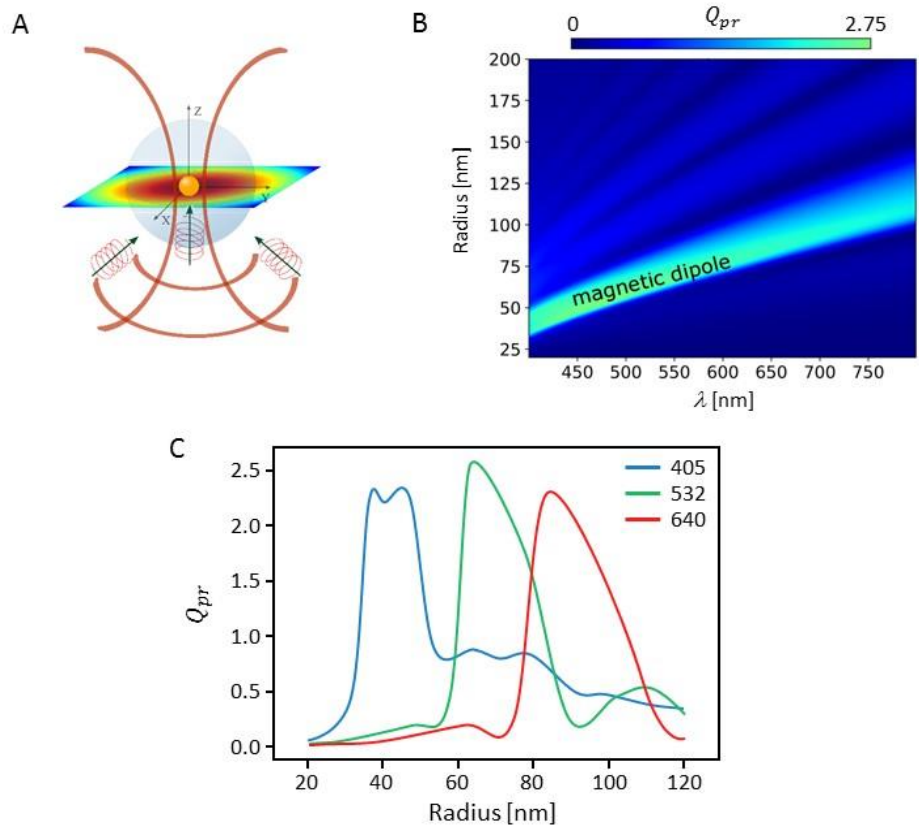

**Figure 2**: Radiation pressure efficiency for a Silicon nanosphere excited by a focused-dipolar laser beam when the particle is located at the focus center. (A) Laser beam scheme; (B) Dependency of radiation pressure efficiency $Q_{pr}$ on the wavelength and the particle radius; (C) $Q_{pr}$ vs. particle radius at three selected wavelengths used in the experiments: 405 nm, 532 nm, and 640 nm.

Next, we tested experimentally the theoretically predicted size-selective optical sorting of Si NPs as depicted in Figure 3A. Briefly, a drop of polydisperse colloidal suspension of Si NPs was placed on a negatively charged glass substrate. Since the NPs are also negatively charged, no spontaneous attachment of NPs to the substrate occurs due to electrostatic repulsion. Optical printing is achieved with a focused laser beam, generating a small region where the axial optical forces are stronger than the repulsion from the substrate[35]. NPs that have the right size to interact strongly with the focused laser beam are able to overcome the electrostatic repulsion and get fixed to the substrate by van der Waals interactions.[35]

The optical printing process is implemented in a home-built optical microscope that combines dark-field imaging in wide field with three linearly polarized CW lasers used for printing at 405 nm, 532 nm and 640 nm focused to their diffraction limit, as schematically represented in Figure 3B. Live analysis of the confocal scattering signal of each laser allows the detection of the printing events. More details about the automated optical printing procedure can be found in our previous reports.[35,41]

Figure 3C shows example dark field images of Si NPs grids printed with light of 405 nm, 532 nm, and 640 nm wavelength. The size selectivity is already evident from the NPs colors. The size of each printed NP was determined using a field emission scanning electron microscope (FE-SEM). Figure 3D shows the size distribution of the NPs printed with each wavelength. The radius distributions present mean values of 43 nm, 59 nm and 73 nm for printing at 405 nm, 532 nm and 640 nm wavelength, respectively. Remarkably, the standard deviation for all distributions is smaller than 9 nm.

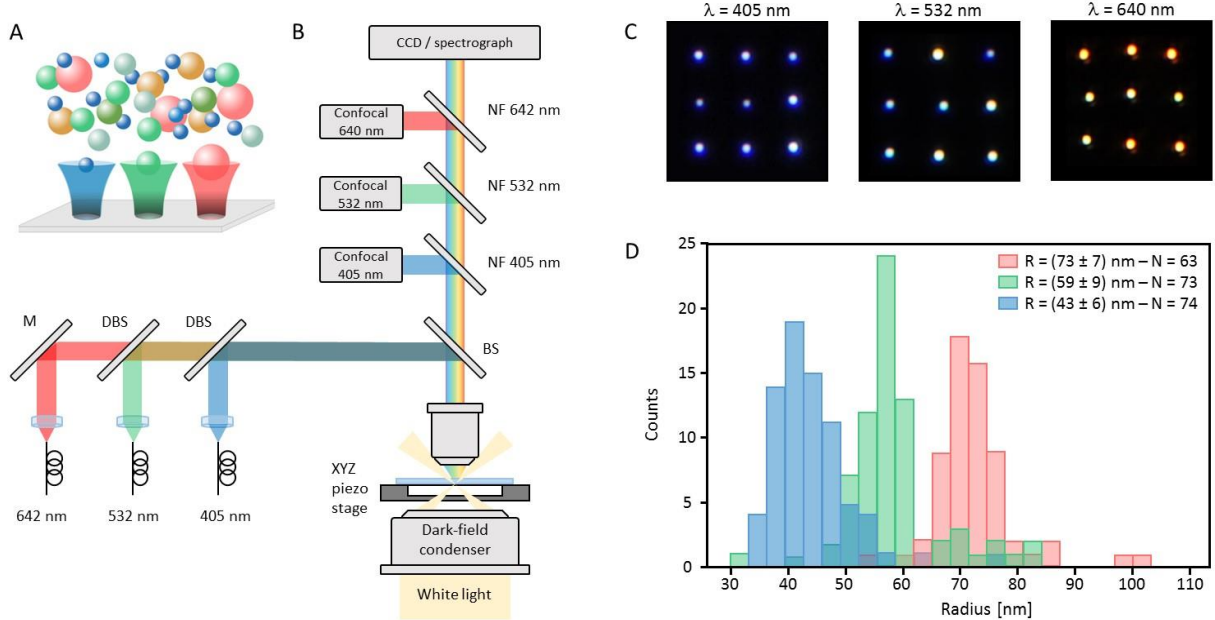

**Figure 3.** Size-selective optical printing of Si NPs. (A) Schematic of the size-selective optical printing process. (B) Sketch of the experimental set-up. (C) Example grids and (D) size distributions of NPs optically printed at 405 nm, 532 nm, and 640 nm.

Using the right laser power is crucial for optical printing accuracy.[35] At lower powers, optical forces become too weak to overcome the electrostatic repulsion and printing events are infrequent. Contrary, at very high powers, strong enough forces are generated over larger regions leading to a faster printing rate but at the expense of accuracy[35]. In the present experiments, laser power is also a key parameter to produce differential printing of NPs with different sizes. Irradiances at the sample were 1.2 MW/cm2, 1.7 MW/cm2 and 1.8 MW/cm2 for the 405 nm, 532 nm and 640 nm beams, respectively. Ideally, low laser powers close to the printing threshold are desirable in order to maximize selectivity and accuracy. Also, lower irradiances are in principle required to print larger particles, as deduced from the similar $Q_{pr}$ for the three wavelengths (Figure 2c) and the scaling of the radiation forces with the geometrical cross section of the particle (Equation 2). However, it must be noted that the proportion of larger NPs in the original suspension is small. Most of the NPs have radii below 50 nm (Figure 1b) and are printed effectively with light of 405 nm (Figure 3D). Consequently, at equal optical force, significantly slower printing rate were observed for larger NPs. In order to obtain faster and practical printing rates of the unabundant larger NPs, higher intensities (and not lower) were used at 532 nm and 640 nm. As expected[35], the faster printing comes at the expense of accuracy, as can be seen in the grids of Figure 3C. Using lower intensities would allow better accuracy and selectivity but, for this particular size distribution of the Si NPs, at the cost of very slow printing rates. It is also interesting to note that the irradiances used to print Si NPs are in the same order of magnitude that the ones used for optically print metallic Au or Ag 60 nm NPs (typically around 1MW/cm$^2$ when printing on resonance).[35]

To further characterize the optical selectivity, correlated dark-field scattering spectroscopy and FE-SEM imaging was performed on individual Si NPs optically printed at the different wavelengths. Figures 4A-C show the scattering spectra, dark field and FE-SEM images of selected Si NPs printed with the 405 nm, 532 nm and 640 nm laser. Scattering spectra of Si NPs printed at 405 nm (R = 43 ± 6 nm) and at 532 nm (R = 59 ± 9 nm) present a single peak, corresponding to the magnetic dipolar resonance. The NPs printed at 640 nm (R = 73 ± 7 nm) present two peaks; the magnetic dipolar resonance shifts towards longer wavelengths, and the electric dipolar resonance appears in the blue region.[53] Figure 4D shows the position of the scattering maximum corresponding to the magnetic dipolar resonance as a function of the radius for 115 Si NPs. By simple inspection, it can be noted that the experimental maximum wavelength is clearly defined by the size of the printed NP. The magnetic dipolar resonance occurs at a wavelength $\lambda = 2Rn$, where $n$ and $R$ are the particle refractive index and radius, respectively.[14,53,54] This tendency is plotted in Figure 4D as a solid grey line. The agreement between experiments and the theoretical description is quite remarkable, and demonstrate that the observed scattering maxima correspond to the magnetic dipolar resonance, as expected for this size range. In addition, the calculated wavelengths for the maximum radiation pressure (same data as Figure 2B) is also shown in Figure 4D as a grey area. In order to selectively print NPs of a given size, the wavelength of the beam should be tuned to the magnetic dipolar resonance.[46] It should be noted that this is valid in the explored range of NPs sizes (10 – 90 nm radius), and specifically when using a tightly focused laser beam. Optical printing with size selectivity is then possible due to the presence of a unique resonance (magnetic) that primarily contributes to the optical force for a given size.

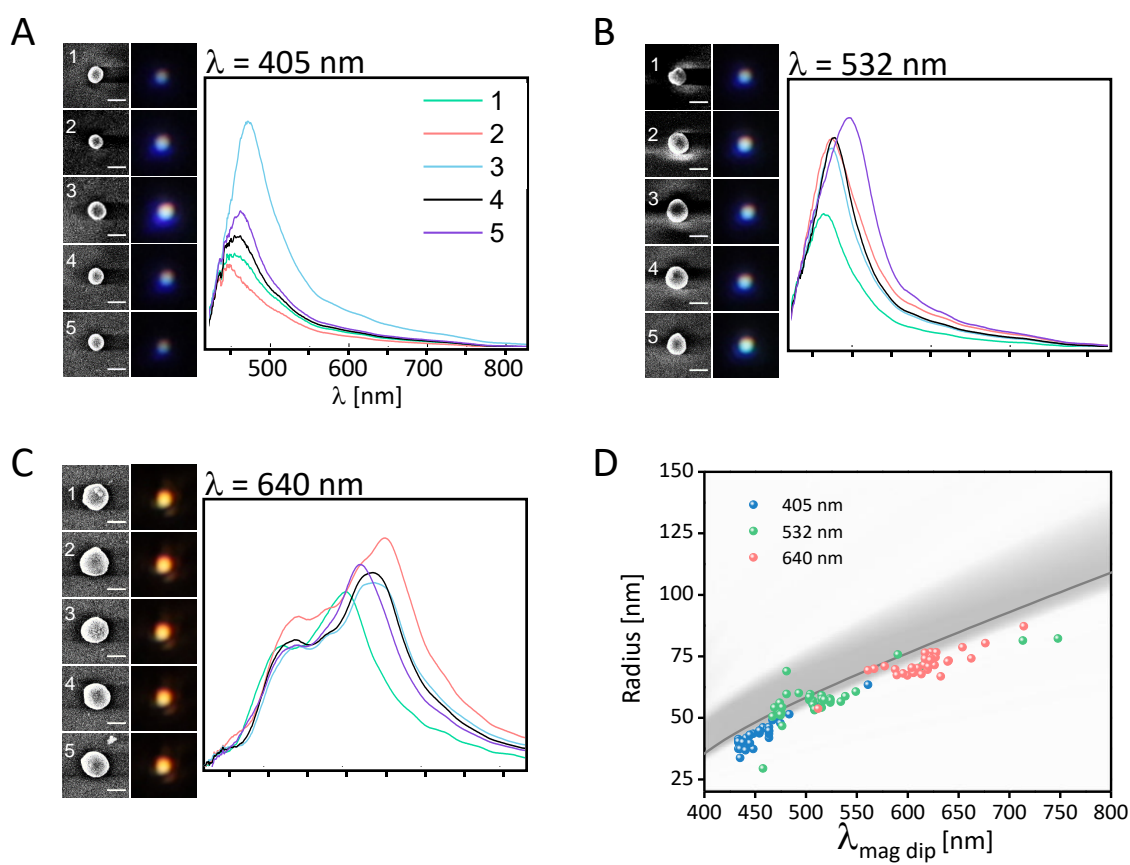

**Figure 4.** Size selectivity and optical properties of the optically printed Si NPs. (A-C) Scattering spectra, dark field and FE-SEM images (scale bar = 100 nm) of selected single Si nanoparticles printed at different wavelengths: (A) 405 nm, (B) 532 nm, (C) 640 nm; (D) Radius vs. scattering maximum of the magnetic resonance for 115 Si NPs printed with the three

wavelengths. The grey solid line is the theoretical prediction $\lambda = 2Rn$. The grey area indicates the maximum radiation pressure vs radius for the magnetic resonance (grey area, same data as in Figure 2B).

In conclusion, differently sized Si NPs can be selected and deposited one by one on a substrate by optical printing from a polydisperse colloid. Starting from a suspension containing NPs with radii ranging from 10 to 100 nm, subpopulations with a radius dispersion below 10 nm can be selectively printed using monochromatic laser light. Tuning the laser wavelength gives control over the average NP size; printing with light of 405 nm, 532 nm, and 640 nm deposits Si NPs with radii of $(43 \pm 6)$ nm, $(59 \pm 9)$ nm and $(73 \pm 7)$ nm, respectively. The NP size as well as the size dispersion of the optically printed NPs is in remarkable agreement to the theoretical predictions of radiation pressure exerted with a tightly focused, monochromatic laser beam. Under these conditions, the radiation pressure on HRI NPs is dominated by the magnetic dipolar resonance, whose size dependency is the basis for the NP size selectivity. This technique enables experimental optical sorting of HRI NPs directly by its optical function, rather than hydrodynamic diameter or density. Future applications include systematic studies correlating optical properties to morphology of a large number of single NPs, and nanofabrication using HRI NPs as building blocks in combination with other colloidal nanomaterials.

ASSOCIATED CONTENT

**Supporting Information**.

Details about theoretical calculations. This material is available free of charge via the Internet at http://pubs.acs.org.

AUTHOR INFORMATION

**Corresponding Author**

*ianinav@conicet.gov.ar; fernando.stefani@cibion.conicet.gov.ar

**Funding Sources**

This project was funded with the support of the following grants: CONICET PIO 13320130100199CO, PICT-2013-0792, PICT- 2014-3729 and a Partner Group of the Max-Planck-Society. We thank also the Spanish Ministerio de Economía y Competitividad (MICINN) and European Regional Development Fund (ERDF) Project FIS2015-69295-C3-3-P, the Basque Dep. de Educación Project PI-2016-1-0041 and the Alexander von Humboldt Foundation for support through a Georg Forster Research Award to FDS. SAM acknowledges the EPSRC Reactive Plasmonics Programme (EP/M013812/1), the Lee-Lucas Chair in Physics, and the Bavarian Solar Technologies go Hybrid (Soltech) programme. J.G. and E.C. acknowledge financial support from the European Commission through a Marie Curie fellowship 797044 Plasmonic Reactor (J.G.) and ERC starting grant 802989 CATALIGHT (E.C.). S.S. acknowledges the financial support for Ph.D. fellowships from Evonik Industries (to L. L.) and from the Fonds der Chemischen Industrie (to M. K.).

**Notes**

The authors declare no competing financial interest.

ACKNOWLEDGMENT

Authors would like to thank Exemys S.R.L. for electronic hardware support, and María Claudia Marchi and Dr. Lía Pietrasanta of the Advanced Microscopy Center of the Faculty of Exact and Natural Sciences of the University of Buenos Aires for the FE-SEM imaging.

**For Table of Contents Use Only**

Manuscript title:

Size-selective optical printing of silicon nanoparticles through their dipolar magnetic resonance

Authors:

Cecilia Zaza, Ianina L. Violi, Julián Gargiulo, Germán Chiarelli, Ludmilla Langolf, Jurij Jakobi, Jorge Olmos, Emiliano Cortes, Matthias König, Stephan Barcikowski, Sebastian Schlücker, Juan José Saenz, Stefan A Maier, Fernando D. Stefani

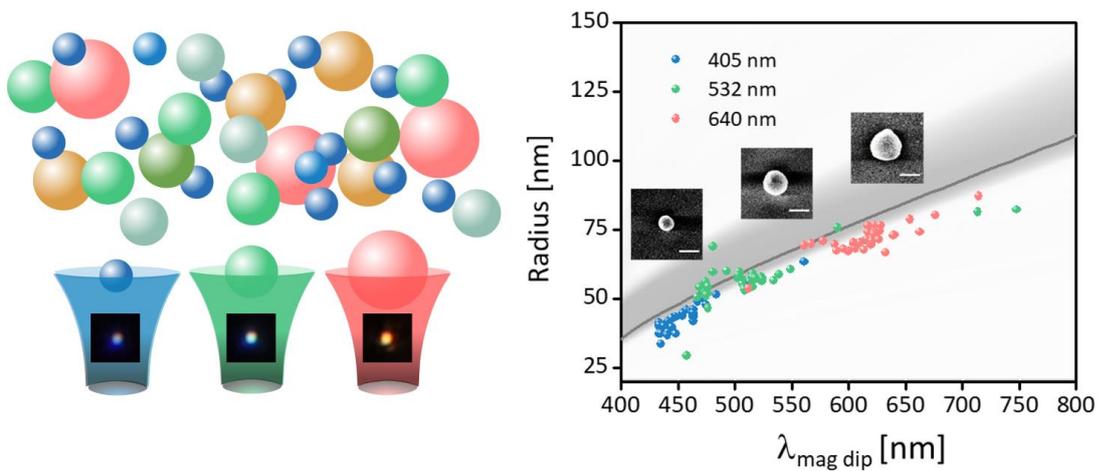